# Transport behavior and electronic structure of phase pure $VO_2$ thin films grown on *c*-plane sapphire under different $O_2$ partial pressure

Salinporn Kittiwatanakul [1], Jude Laverock [2], Dave Newby Jr.[2], Kevin E. Smith [2], Stuart A. Wolf [1,3], Jiwei Lu [3]

[1] Department of Physics, University of Virginia
[2] Department of Physics, Boston University
[3] Department of Materials Science and Engineering, University of Virginia

We grew highly textured phase pure $VO_2$ thin films on *c*-plane $Al_2O_3$ substrates with different oxygen partial pressure. X-ray absorption and photoemission spectroscopy confirm the identical valence state of vanadium ions despite the different oxygen pressure during the deposition. As the $O_2$ flow rate increases, the [010] lattice parameter for monoclinic $VO_2$ was reduced and coincidently distinctive changes in the metal-semiconductor transition (MST) and transport behaviors were observed despite the identical valence state of vanadium in these samples. We discuss the effect of the oxygen partial pressure on the monoclinic structure and electronic structure of $VO_2$, and consequently the MST.

Vanadium oxides have been of considerable interest in the past due to the metal insulator transition (MIT) that several of the oxides exhibit [1,2]. In particular, vanadium dioxide ($VO_2$) exhibits a metal semiconductor transition (MST) at 340 K, yielding an abrupt change in the transport and optical properties [3]. Thanks to the improvements in the quality of $VO_2$ thin films, there has been a growing emphasis on exploring logic and memory device applications of this intriguing phenomenon. Recent efforts on a Mott field effect transistor (MottFET) have demonstrated that $VO_2$ can be employed as the channel in a MottFET, and the applied external gate voltage switches the channel resistance between the insulating OFF state and the metallic ON state [4-6]. There are still several challenges for these so called "Mott devices" [7], among which is a concern over the repeatability/reproducibility of the Mott transition because the properties of these materials are sensitive to oxygen vacancy concentration, strain, and other factors. Understanding the role of defects and developing a suitable growth technique to maintain reproducibility is thus required to realize the potential applications of Mott oxides.

The growth and the physical properties of vanadium dioxide thin films have been explored by various deposition techniques such as electron beam evaporation [8-10], reactive sputtering [10-15], sol-gel process [16,17], chemical vapor deposition [17], atomic layer deposition [18], and pulsed laser ablation [19,20]. Though it was relatively straightforward to achieve a single phase $VO_2$, the MST behaviors, i.e. the transition temperature and the magnitude of resistivity ratio, lacked consistency in the literature. To discern the effect of oxygen content and the valence state of vanadium, we synthesized phase pure $VO_2$ films under different oxygen flow rates. Soft X-ray absorption (XAS) and photoemission (XPS) spectroscopy were used to characterize the thin films and investigate the valence state of the vanadium ion in these films. Despite the fact that the films are single phase, each sample has very distinctive transport properties. A systematic characterization of the electrical properties of these films was carried out using Hall devices, and we will discuss the correlation between the oxygen content, the electronic structure and the MST.

Reactive Bias Target Ion Beam Deposition (RBTIBD) was used [21]. For this study, the main chamber was pumped down to a base pressure of $\sim 5 \times 10^{-8}$ Torr. Vanadium metal target (99.99% purity) was sputtered in a mixture of Ar and $O_2$, and was water-cooled during deposition. The stage heater was ramped to 450 ºC and allowed to stabilize for 45 min. The ion source was started with 10 sccm Ar gas flow and 5 V on the cathode, and 70 sccm Ar gas flow and 40 V on the anode. The substrate was gently pre-cleaned using Ar ions, and the vanadium target was sputter-cleaned by applying pulsed dc bias. The frequency was set at 71.43 kHz with a 3 μs positive duty cycle. For the oxygen source, an Ar/$O_2$ 80/20 mixture was introduced with the flow rate set to 4.5-6.0 sccm. The total processing pressure can be obtained by adding all the gas flowing through cathode (10 sccm Ar), anode (70 sccm Ar), and the additional Ar/$O_2$ 80/20 mixture (4.5-6.0 sccm); the total processing pressure was nearly the same for all depositions, ~ 1.2 mTorr during each deposition. The growth time was 37.5 min.

The film thickness was determined by X-ray reflectivity (Smart-lab, Rigaku Inc.), and the thickness of each film is shown in table I. Out-of-plane and in-plane X-ray diffraction (XRD) scans were also performed to confirm the phase composition, and to determine the lattice parameter of the $VO_2$ films. Atomic Force Microscopy (Cypher®, Asylum Research) and Gwyddion software were used to characterize the surface morphology. AFM has revealed that the lower oxygen flow rate sample has smaller grain and a smoother surface compared to the higher oxygen flow rate sample, i.e. the 4.5 sccm sample has a root-mean-square (RMS) roughness of 0.5 nm, while the 6.0 sccm sample has a RMS roughness of 1.8 nm. There are no cracks or pinholes observed on a large 20 x 20 μm² area. Soft X-ray absorption and photoemission were performed at Beamline X1B of the National Synchrotron Light Source, Brookhaven National Laboratory. XAS measurements were made at a resolution of 0.3 eV in both total electron yield (TEY) and total fluorescent yield (TFY) modes, which have probing depths of 10 nm and 100 nm respectively. XPS measurements were performed with a Scienta-100 hemispherical analyzer, with energy resolution of 0.4 eV for the V $2p$ / O $1s$ core level (h$\upsilon$ = 900 eV) and 0.7 eV for the valence band measurements (h$\upsilon$ = 700 eV). Resonant photoemission spectroscopy (RPES) measurements, in which the incident photon energy is tuned to a feature of the absorption spectrum, were performed across the V $L_{3,2}$-edge (h$\upsilon$ = 510 – 525 eV) in order to resonantly enhance the photoelectron transition rate of the V $3d$ electrons.

To characterize the transport behavior of the films, photolithography and reactive ion etching (RIE) were used to fabricate Hall bars for 4-point measurements. The Ohmic contact was 100 nm Au / 10 nm Ti, deposited by electron beam evaporation. The temperature dependence of the dc resistivity was measured using a Versa-lab system (Quantum Design) from 300 to 400 K with a heating/cooling rate of 2 K/min. The *dc* resistivity was then calculated according to the device geometry and the thickness of the film. The Hall effect was measured using a physical property measurement system (Quantum Design PPMS 6000) from 300 to 400 K with a 10 K increment, sweeping magnetic field from -7 to 7 T at each temperature. The Hall effect measurement was then extended from 200 to 400 K for the 4.5 sccm and 6.0 sccm samples.

XRD scans showed that the VO$_2$ films deposited on the c-plane Al$_2$O$_3$ single-crystal substrate were single phase and highly textured. The out-of-plane $2\theta$ scan [Fig. 1(a)] showed that the (020) peak in the VO$_2$ diffraction was coupled to the (0006) peak of the Al$_2$O$_3$ substrate, and the VO$_2$ peak was the only peak detected in a wide-range $2\theta$ scans (not shown here), which indicated no secondary phases in all the films, despite the fact that they were deposited under various O$_2$ atmospheres. The (020) texture for VO$_2$ was typical when the template is *c*-plane sapphire [22]. There are clear Kiessig fringes in the out-of-plane $\theta$-$2\theta$ scan observed in all samples, indicating high crystallinity with smooth interfaces, uniform thickness and low defect density.

The effect of oxygen flow rate on the lattice parameter '*b*' is shown in Fig. 1(b). From the values of $2\theta$, we estimated the lattice parameters for the (010) reflection, i.e. *b*, of monoclinic VO$_2$ for different films deposited under various Ar/O$_2$ mixture flow rates. Fig. 1(a) shows that as the Ar/O$_2$ flow rate increases, the diffraction peak for the (020) reflection shifted to larger $2\theta$, resulting in a smaller lattice parameter. In comparison, the film deposited using a 6.0 sccm Ar/O$_2$ flow rate has a smaller *b*-spacing ([010] lattice parameter) compared to the film deposited at the 4.5 sccm flow rate. The [010] lattice parameter of these single phase highly textured thin films are smaller than those of bulk VO$_2$ which is 4.517 Å [23].

To identify the valence state of the vanadium ion, soft X-ray absorption spectra were obtained [24]. For clarity, only spectra from the 4.5 sccm and 6.0 sccm samples are shown in Fig. 2. Fig. 2(a) shows the V $L_{3,2}$-edge and the O *K*-edge absorption spectrum. There is no indication of the mixed valence states of V despite the difference in the oxygen flow rate and the lattice parameter, both in TEY (more surface sensitive) and TFY (more bulk sensitive) modes. Different V charge states have quite different multiplet structures in the V $L_{3,2}$-edge absorption spectrum, the most obvious difference being in their peak location, a change that we do not observe. The valence state of vanadium is also determined by V $2p$ XPS as shown in Fig. 2(b). The two structures are V $2p_{1/2}$ and $2p_{3/2}$ core levels. These are very sensitive to the V charge state, with the peak feature shifting by around 2 eV from $V^{3+}$ to $V^{5+}$ (as indicated). According to Silversmit et al. [25], the peak energies in Fig 2(b) are consistent with the $V^{4+}$ oxidation state, i.e. to $VO_2$. There is no shift in the V $2p$ core levels between the two extreme end-members, indicating the valence state of vanadium cations does not change. XPS is surface sensitive; thus these results indicate that at least the surface is $V^{4+}$ across the series. The absence of any shift in the V $L_{3,2}$-edge TFY [Fig. 2(a)], a bulk sensitive mode, suggest that the valence state of vanadium is also $V^{4+}$ across the bulk of the films.

It is worth noting that there are two subtle features (circled) in the absorption spectrum that show some differences between the two end-members. Firstly, the pre-edge of the V $L_3$-edge, at around 514 eV, extra weight was observed in the 4.5 sccm sample. These states correspond to unoccupied states available at lower energy, indicating that the band gap has closed above the Fermi level for the 4.5 sccm sample. However, as shown in the O $K$-edge absorption [Fig. 2(a)], there is only a very weak change in the onset energy between the two samples. Therefore, the extra states available at lower energy are associated with pure V $3d$ states, i.e. with the V-V (along the $c$-axis) bonds. Secondly, there is an evolution in the crystal-field splitting of the $t_{2g}$ and $e_g$ states, shown in the O $K$-edge spectrum [Fig. 2(a)]. This is most likely due to the evolution in the structural distortion of the VO$_6$ octahedra associated with the change in the *b*-axis lattice parameter. The 6.0 sccm sample has a smaller *b*-axis lattice parameter than 4.5 sccm. This means that the V 3*d* states feel the effects of the crystal field more strongly, and the crystal field splitting increases. In other words, the 6.0 sccm sample should have a larger separation between *t*$_{2g}$ and *e*$_g$ states, consistent with XAS results.

The resistivity as a function of temperature is shown in Fig. 3(a). There are very distinguishable differences in the transport properties for the various samples, despite the fact that all of the samples are phase pure VO$_2$ with the same valence state for the vanadium cations. The insulating phase of the lower Ar/O$_2$ flow rate samples is more conductive than that of the higher Ar/O$_2$ flow rate samples, while the resistivity of all samples in the metallic phase are at about the same value; hence the lower Ar/O$_2$ flow rate samples yield much smaller change in resistivity across the MST. The 6.0 sccm sample has a highest change in resistivity, about 3 orders of magnitude, while the change in resistivity of the 4.5 sccm sample drops down to just below an order of magnitude.

The resistivity measurements of all samples also show a hysteresis loop around T$_{MST}$ of each sample, which gets smaller as the Ar/O$_2$ flow rate increases. The Metal Semiconductor Transition Temperature (T$_{MST}$) of each sample was extracted from the derivative of the logarithm of the resistivity, that is defined as

$$T_{MST} = \frac{T_{up} + T_{down}}{2}$$

when $T_{up} = T$ where $\frac{d(\log \rho_{up})}{dT}$ is at a minimum, and $T_{down} = T$ where $\frac{d(\log \rho_{down})}{dT}$ is at a minimum. $\rho_{up}$ is the resistivity of the up-sweep (increasing temperature from 300 to 400 K), and $\rho_{down}$ is the resistivity of the down-sweep (decreasing temperature from 400 to 300 K). Extracting the results from resistivity measurements and XRD results clearly show that as the (010) spacing increases, the $T_{MST}$ of the sample also decreases toward room temperature as shown in Fig. 1(c). The shift in the $T_{MST}$ and the drastic change in the transport properties are most likely due to the change in (010) spacing and the distortion introduced by different $Ar/O_2$ flow rate. The epitaxial strain in $VO_2$ grown on c-plane $Al_2O_3$ substrates was fully relaxed at the thickness of 25 nm due to the large lattice mismatch between $VO_2$ and $Al_2O_3$ [26], hence we may conclude that the epitaxial strain was similar in 4.5 to 6 sccm samples despite the various thicknesses. According to the unit cell volume conservation, as the b-spacing increases (larger (010) spacing), the c-axis is expected to be shorten proportionally, and $T_{MST}$ decreases accordingly, which is good agreement with previous report by Y. Muraoka [27]. Even though the unit cell volume conservation may not be true, especially for the thin film; however, the relationship of $T_{MST}$ and oxygen partial pressure is in a good agreement with a previous report [15]. The effect of oxygen partial pressure on the out-of-plane lattice parameter is also similar with Kaushal and Kuar's result when they deposited $(WO_3)_{1-x}(VO_2)_x$ nanocomposite thin films by pulsed laser deposition [28].

The strain to cause the change in the spacing was not pre-determined by the film thickness. The effect of film thickness on transport behavior has been reported previously [26]. The transition temperature was shifted from lower temperatures to 340 K with the increase in the thickness, which was attributed to the strain relaxation as a function of film thickness. We have observed the opposite trend as summarized in Table I, which mean that the $T_{MST}$ reported here is more likely the effect of the oxygen partial pressure than the film thickness.

The Hall effect was measured on all samples from 300 to 400 K with a 10 K increment, sweeping magnetic field from -7 to 7 T at each temperature, and the extracted carrier density and carrier mobility are shown in table I. There exists an abrupt change in the carrier density around the transition temperature, while the mobility seems to be constant before and after the transition. In the vicinity of the transition temperature, for the 5.0 and 6.0 sccm samples, there occurs a maximum of the mobility, which is approximately 10-100 times larger than the mobility before and after the transition. Recently, Ruzmetov et al. [29] have reported the same trend of mobility for $VO_2$ thin films grown on *c*-plane $Al_2O_3$ substrates. Instead of the maximum, Ruzmetov et al. found that around the MST, there was a minimum, which was ten times smaller than the mobility both below and above $T_{MST}$. Unlike the higher flow rate samples, the 4.5 sccm sample has a very high carrier density, which does not have a large change around the transition temperature, while the mobility remains extremely low. The mobility of the 5.5 sccm sample looks more like the 4.5 sccm sample, since there is no maximum around $T_{MST}$. The majority carriers of all samples are electrons above room temperature.

Resonant photoemission spectra (resonant with the V $L_3$ edge) are shown in Fig. 3(b). The RPES enhances the photoemission transition rate of those states that the incident photons are resonant with, in this case the V 3*d* states. The RPES results show the V 3*d* band more clearly across the transition. On both resonances ($L_3$ and $L_2$), the insulating gap is found to be much smaller for the 4.5 sccm sample, with the 6.0 sccm sample having a larger gap. The other two samples lie approximately in between, although the 5.0 sccm sample has a larger gap than the 5.5 sccm sample. It shows that the V 3*d* states were found to extend much closer to the Fermi level in the insulating phase of 4.5 sccm compared with 6.0 sccm. The electronic structure observed near the Fermi level was expected in the light of the transport behaviors shown in Fig. 3(a), where the insulating phase of the 4.5 sccm sample is much more conductive than that of the 6.0 sccm sample. The soft X-ray spectroscopy results show that the insulating gap closes symmetrically (i.e. both occupied and unoccupied states close) for the 4.5 sccm sample. The states involved in closing the gap are most likely V-V bonding electrons, since the O partial density of states shows only weak evolution with flow rate. It is likely that oxygen-deficient growth condition probably leads to the introduction of interstitial vanadium ion (do not contribute to the electronic structure) [30], hence increase the carrier density, and reduce the resistivity in the semiconductor state (RT). The interstitial defect results in the increasing [010]-lattice parameter. The *c*-axis of the oxygen-deficient grown sample thus became smaller, and hence had a smaller $T_{MST}$, as previously reported on epitaxial $VO_2$ deposited on $TiO_2$ [27]. On the other hand, oxygen-rich growth leads to V vacancies, reducing the [010]-lattice parameter. It is consistent with the transport behaviors if the interstitial V doping level is dilute (below our limits of detection) or the interstitial V only distorts the lattice, and does not donate electrons like $V^{4+}$ cations on the lattice sites. Both the defects and the change in *b*-spacing directly contribute to the conductivity, especially in the insulating phase, but the effects seen in the electronic structure, i.e. closing of the gap, are more related with the strain. The results are consistent with interstitial V, however to properly study other type of defects such as the oxygen vacancies, other techniques would be needed.

Hall mobility measured at room temperature did not significantly change throughout the series, which agree with the measurements conducted by Kwan *et al.* [31], so the resistivity is inversely proportional to the carrier density. In the semiconductor state, the crystal distortion reduces the stabilization, thus these extra carriers (electrons at RT) tend to lower the $\pi^*$ level [30], and may screen the electron correlation energy of $d_\parallel$ electrons [32], hence reduce the gap of $d_\parallel$ bands. In addition, the population of these thermally excited electrons increases in temperature, as shown in carrier density extracted from the Hall effect measurements, thus the effect enhances around $T_{MST}$, and narrow of the band gap of $VO_2$, resulting in a less sharp transition as shown in the resistivity measurement in Fig. 3(a).

Despite the fact that the samples are single phase $VO_2$ and the valence state of V does not change much through the series, with predominantly $V^{4+}$ in each sample, the transport properties of these samples are varied due to the different $Ar/O_2$ flow rates during the deposition. The Hall effect measurements reveal very significant difference in carrier densities between higher and lower oxygen flow rate samples; the resistance as a function of temperature measurements also shows a shift of $T_{MST}$, and that the insulating phase of lower $Ar/O_2$ flow rate samples are much more conductive. It was likely due to the strain that was caused by the different types of defects associated with the oxygen growth environments. The soft X-ray spectroscopy results show subtle evidence for the crystal distortions reflected by the different crystal-field splitting between the 4.5 and 6.0 sccm samples. Our findings suggest that the repeatability/reproducibility of the Mott transition in $VO_2$ can be obtained when one discerns the effect of the growth condition on the film strain and the electronic structures.

**Acknowledgement:**

S. K, S.A. W and J.W. L are grateful to the support from Nanoelectronics Research Initiative and VMEC. The Boston University program is supported in part by the US Department of Energy under Contract No. DE-FG02-98ER45680. The NSLS, Brookhaven, is supported by the US Department of Energy under Contract No. DE-AC02-98CH10886.

**Figure Captions**

Fig. 1 (a) Room temperature XRD data of $Al_2O_3$ (0006) and $VO_2$ (020). (b) The effect of $O_2$ flow rate on $b$-spacing. (c) The effect of $b$-spacing on the transition temperature.

Fig. 2 (a) V $L_{3,2}$-edge and O $K$-edge XAS, TFY is more bulk sensitive, and TEY is more surface sensitive. (b) V $2p$ XPS.

Fig. 3 (a) dc resistivity as a function of temperature. (b) V $L_3$-edge resonant photoemission. Dash lines are HT (metallic), solid lines are RT (insulating).

**Fig. 1**

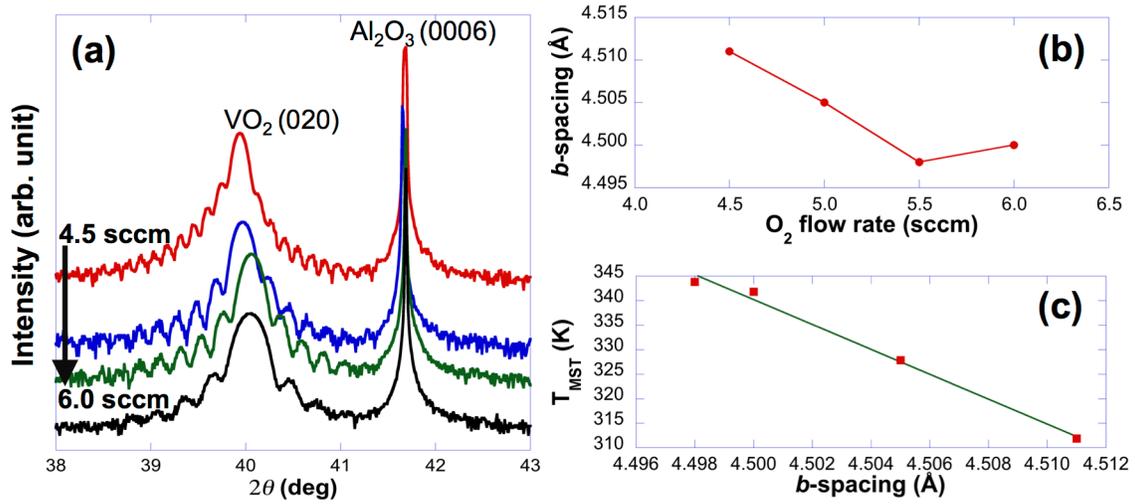

**Fig. 2**

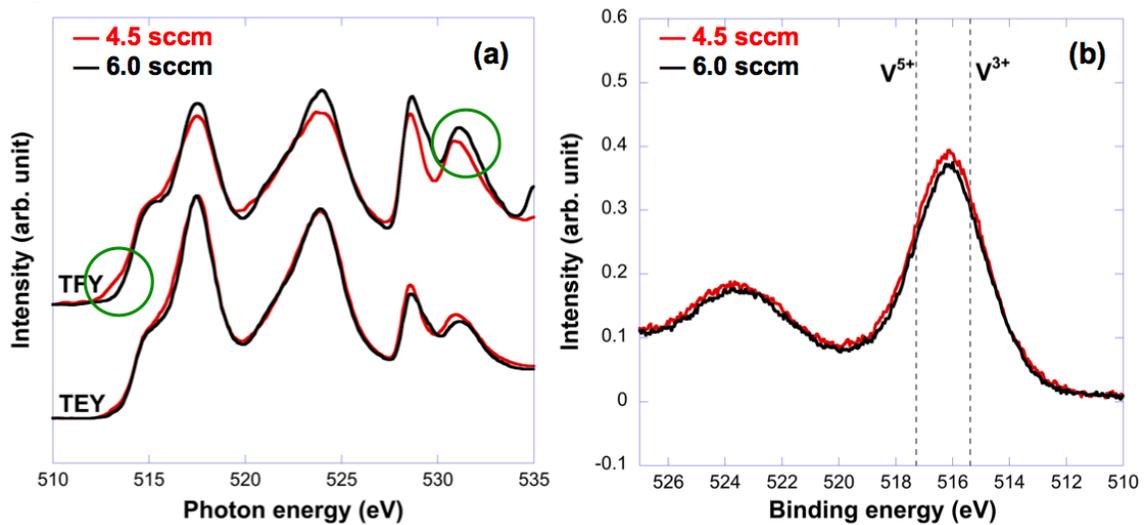

Fig. 3

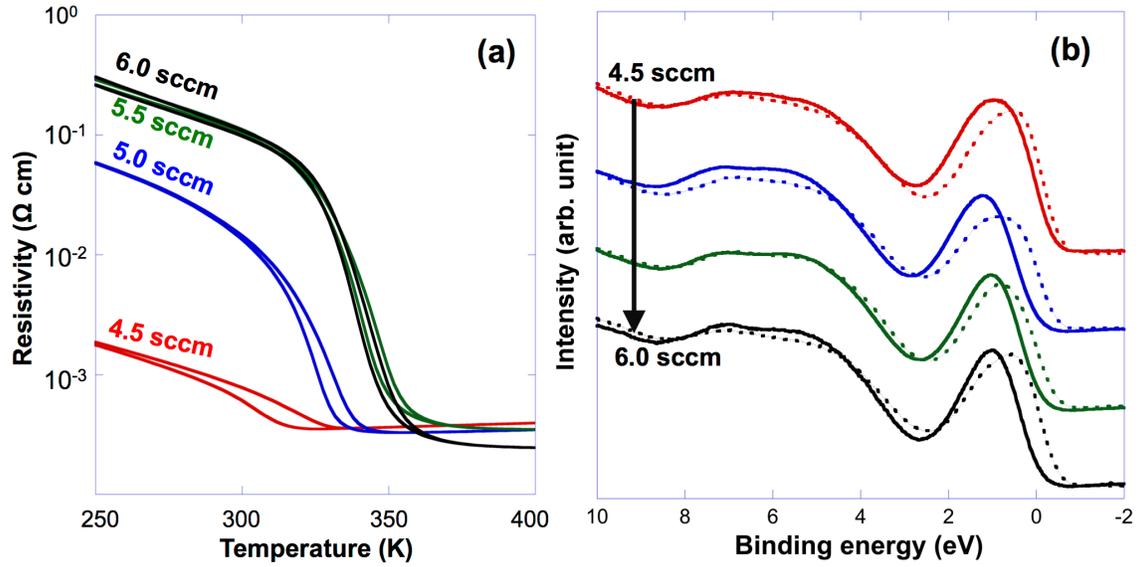

Table I. Electrical transport properties of the 4 samples with different $O_2$ flow rate

|  | 4.5 sccm | 5.0 sccm | 5.5 sccm | 6.0 sccm |
|---|---|---|---|---|
| **Thickness (nm)** | 66 | 50 | 45 | 35 |
| $T_{MST}$ **(K)** | 311.9 | 327.8 | 343.8 | 341.8 |
| $R_{300K}/R_{400K}$ | ~2.0 | ~41.9 | ~318.5 | ~440.5 |
| $n_{300K}$ **(×10²⁰ cm⁻³)** | ~753 | ~1.49 | ~2.67 | ~0.24 |
| $n_{400K}$ **(×10²⁰ cm⁻³)** | ~1250 | ~6130 | ~4140 | ~104 |
| $\mu_{300K}$ **(cm²/Vs)** | 0.05 | 0.80 | 0.20 | 0.70 |
| $\mu_{400K}$ **(cm²/Vs)** | 0.06 | 0.05 | 0.04 | 0.80 |